\documentclass{svmult}
\usepackage{pstricks}
\usepackage{pst-plot,pst-xkey}
\usepackage{multido}
\makeatletter
\newcommand{\radialplot}{\pst@def{ScalePoints}<%
  /y ED /x ED
  counttomark dup dup cvi eq not { exch pop } if
  /m exch def /n m 2 div cvi def
  n { /RAD exch def  
      dup            
      cos exch sin RAD mul y mul m 1 roll
      RAD mul x mul m 1 roll
      /m m 2 sub def } repeat>}
\def\reflineA{
  90 1 162 1  234 1 306 1 18 1 
}%
\def\reflineB{
  90 0.5 162 0.5  234 0.5 306 0.5 18 0.5 
}%
\makeatother
\newcommand{\plotaxesetc}{
  \multido{\rA=0.0+72.0}{5}{%
    \rput{\rA}(0,0){\psline[linewidth=0.5pt]{->}(0,0)(0,1.1)}
  }
  \SpecialCoor
  \rput[b](3.4;90){PMEMD}
  \rput[rb](3.4;162){EXX}
  \rput[t](3.4;234){IMD}
  \rput[t](3.4;306){Oak3D}
  \rput[lb](3.4;18){TRATS}
  \rput[lb](3.1;89){\small 1}
  \rput[lb](1.55;89){\small 0.5}
  \radialplot
  \listplot[linecolor=black,linewidth=1.0pt,plotstyle=polygon,linestyle=dotted]{\reflineA}
  \listplot[linecolor=black,linewidth=1.0pt,plotstyle=polygon,linestyle=dotted]{\reflineB}}
\newcommand{\plotdatafile}[3]{%
  \readdata{\data}{#1}
  \listplot[linecolor=#2,linewidth=2pt,plotstyle=polygon,linestyle=solid,
    showpoints=true,dotstyle=#3]{\data}}
\newcommand{\plotaxesetcnum}[5]{
  \multido{\rA=0.0+72.0}{5}{%
    \rput{\rA}(0,0){\psline[linewidth=0.5pt]{->}(0,0)(0,1.1)}
  }
  \SpecialCoor
  \rput[b](3.4;90){#1}
  \rput[rb](3.4;162){#2}
  \rput[t](3.4;234){#3}
  \rput[t](3.4;306){#4}
  \rput[lb](3.4;18){#5}
  \rput[lb](3.1;89){\small 1}
  \rput[lb](1.55;89){\small 0.5}
  \radialplot
  \listplot[linecolor=black,linewidth=1.0pt,plotstyle=polygon,linestyle=dotted]{\reflineA}
  \listplot[linecolor=black,linewidth=1.0pt,plotstyle=polygon,linestyle=dotted]{\reflineB}}
\newcommand{\plotdatafileopen}[3]{%
  \readdata{\data}{#1}
  \listplot[linecolor=#2,linewidth=2pt,plotstyle=line,linestyle=solid,
    showpoints=true,dotstyle=#3]{\data}}
\usepackage[dvips]{graphicx}    
\usepackage{color}
\usepackage{url}
\usepackage{ragged2e}

\begin{document}
\title*{RZBENCH: Performance evaluation of current HPC architectures
        using low-level and application benchmarks}
\titlerunning{RZBENCH}
\author{Georg Hager\and Holger Stengel\and Thomas Zeiser\and Gerhard Wellein}
\authorrunning{G.~Hager, H.~Stengel, T.~Zeiser, G.~Wellein}
\institute{Regionales Rechenzentrum Erlangen (RRZE), Martensstr.~1, D-91058 Erlangen, Germany}
\maketitle

\begin{abstract}
RZBENCH is a benchmark suite that was specifically developed to
reflect the requirements of scientific supercomputer users at the
University of Erlangen-Nuremberg (FAU)\@. It comprises a number of application
and low-level codes under a common build infrastructure that fosters
maintainability and expandability. This paper reviews the structure of
the suite and briefly introduces the most relevant benchmarks. In
addition, some widely known standard benchmark codes are reviewed in
order to emphasize the need for a critical review of often-cited
performance results. Benchmark data is presented for the
HLRB-II at LRZ Munich and a local InfiniBand Woodcrest cluster as well
as two uncommon system architectures: A bandwidth-optimized InfiniBand
cluster based on single socket nodes (``Port Townsend'') and an early
version of Sun's highly threaded T2 architecture (``Niagara 2'')\@.
\end{abstract}

\section{Introduction}

Benchmark rankings are of premier importance in High Performance
Computing. Decisions about future procurements are mostly based on
results obtained by benchmarking early access systems. Often,
standardized suites like SPEC \cite{spec} or the NAS parallel
benchmarks (NPB) \cite{npb} are used because the results are publicly
available. The downside is that the mixture of requirements to run the
standard benchmarks fast is not guaranteed to be in line with the
needs of the local users.  Even worse, compiler vendors go to great
lengths to make their compilers produce tailor-made machine code for
well-known code constellations.  This does not reflect a real user
situation.

For those reasons, the application benchmarks contained in the RZBENCH
suite are for the most part widely used by scientists at FAU\@.  They
have been adapted to fit into the build framework and produce
comprehensible performance numbers for a fixed set of inputs.  A
central customized makefile provides all the necessary information
like names of compilers, paths to libraries etc. After building the
suite, customizable run scrips provide a streamlined user interface by
which all required parameters (e.\,g., numbers of threads/processes
and others) can be specified. Where numerical accuracy is an issue,
mechanisms for correctness checking have been employed. Output data is
produced in ``raw'' and ``cooked'' formats, the latter as a mere
higher-is-better performance number and the former as the full output
of the application.  The cooked peformance data can then easily be
post-processed by scripts and fed into plotting tools or spreadsheets.

The suite contains codes from a wide variety of application areas and
uses all of the languages and parallelization methods that are
important in HPC: C, C++, Fortran 77, Fortran 90, MPI, and OpenMP\@.

\section{Benchmark systems}

All state-of the art HPC systems are nowadays based on dual-core and
quad-core processor chips. In this analysis the focus is on standard
dual-core chips such as the Intel Montecito and Intel Woodcrest/Conroe
processor series.  The Intel Clovertown quad-core series is of no
interest here, since it implements two complelety separate dual-core
chips put on the same carrier.  We compare those standard technologies
with a new architecture, the Sun UltraSPARC T2 (codenamed ``Niagara
2''), which might be a first glance at potential future chip designs:
A highly threaded server-on-a-chip using many ``simple''
cores which run at low clock speed but support a large number of
threads.

\begin{table}[htbp]
\caption{\label{TabSocketProp} Specifications for the different compute 
nodes, sorted according to single core, single socket and single node
properties. The L2 cache sizes marked in {\bf bold face} refer to
shared on-chip caches, otherwise all caches are local to each
core.}
\renewcommand{\arraystretch}{1.5}
\begin{center}
\begin{tabular}[t]{l|cc|ccc|cc|c}
                                 &\multicolumn{2}{c|}{\bf Core} &\multicolumn{3}{c|}{\bf Cache}&  \multicolumn{2}{c|}{\bf Socket} & {\bf Node}\\ \hline
\multicolumn{1}{c|}{\bf Platform} & \parbox[b]{1.0cm}{\centering \rule{0pt}{1.2em}Clock\\ GHz} & \parbox[b]{1.5cm}{\centering Peak\\ GFlop/s} & \parbox[b]{1.0cm}{\centering L1\\ kB} &  \parbox[b]{1.0cm}{\centering L2\\ MB}  &  \parbox[b]{1.0cm}{\centering L3\\ MB}  & \parbox[b]{1.0cm}{\centering \# of \\ cores}  & \parbox[b]{1.2cm}{\centering Bandw.\\ GB/s}  & \parbox[b]{1.0cm}{\centering \# of\\ sockets} \\ \hline
%
%
HLRB II       & 1.6     & 6.4 & 16  & 0.25    & 9  & 2 & 8.5  & 1-256 \\ \hline
Woodcrest     & 3.0     & 12.0 & 64 & {\bf 4} & -- & 2 & 10.6  & 2 \\
Conroe        & 2.66    & 10.6 & 64 & {\bf 4} & -- & 2 & 8.5   & 1 \\ \hline
Niagara2      & 1.4     & 1.4  & 8 & {\bf 4} & -- & 8 & 42R+21W & 1 \\
\end{tabular}
\end{center}
\end{table}

\subsection{HLRB II -- SGI Altix 4700}

The SGI Altix 4700 system at LRZ Munich comprises 9728 Intel Itanium2
processor cores integrated into the SGI NUMALink4 network.  It is
configured as 19 ccNUMA nodes each holding 512 cores and a total of
2\,Tbyte of shared memory per partition. The 13 standard nodes
are equipped with a single socket per memory channel, while in the
six ``high density'' nodes two sockets, i.e.\ four cores, have
to share a single memory channel. Table~\ref{TabSocketProp} presents
the single core specifications of the Intel Itanium2 processor used
for HLRB II\@. A striking feature of this processor is its large
on-chip L3 cache of 9\,Mbyte per core.  A more detailed discussion of
the Intel Itanium2 architecture is presented in Ref.~\cite{deserno}\@.

\begin{figure}[tbp]
\centerline{%
\includegraphics*[width=0.9\textwidth]{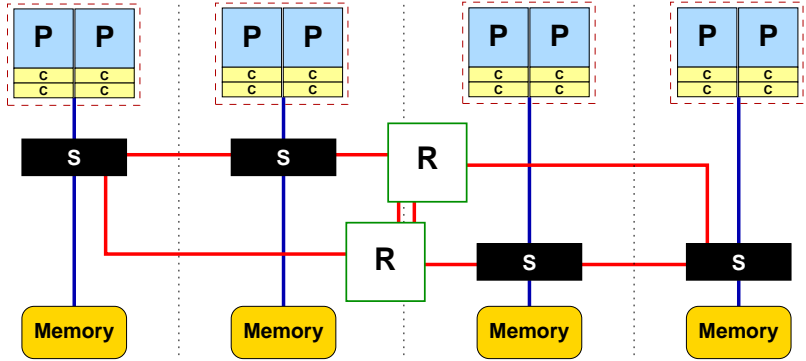}}
\caption{\label{fig:altix}Possible layout of a small SGI Altix system.
        The lines connecting routers (``R'') and SHUB chips (``S'') are
        NUMALink4 connections with a theoretical bandwidth of 3.2\,GB/sec
        per direction.}
\end{figure}

The NUMALink4 network provides a high bandwidth (3.2\,Gbyte/s per
direction and link), low latency (MPI latency can be less than 2
$\mu$s) communication network (see Fig.~\ref{fig:altix} for a possible
network topology in a small Altix system)\@.
However, the network
topology implemented does not allow to keep the bi-sectional bandwidth
constant within the system. Even the nominal bisection bandwidth per
socket (0.8\,Gbyte/s per direction) in a single standard node (256
sockets) falls short of a single point to point connection by a
factor of four. Connecting the nodes with a 2D torus NUMALink
topology, things get even worse. For a more detailed picture of the
current network topology status we refer to
Ref.~\cite{hlrb-topology}\@.

All measurements presented were done within a single standard node.

\subsection{Woodcrest -- InfiniBand cluster}

\begin{figure}[tbp]
\centerline{%
\includegraphics*[width=0.6\textwidth]{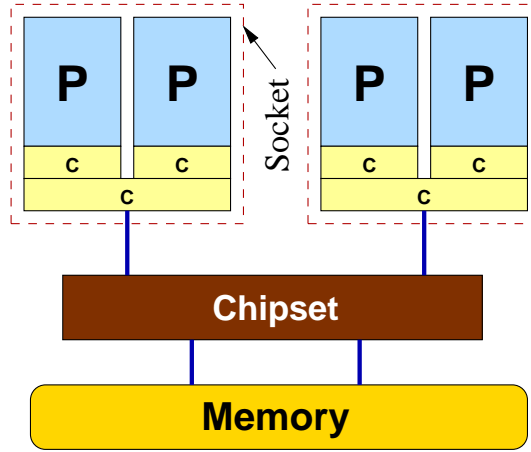}}
\caption{\label{fig:woodcrest}Layout of a single Woodcrest-based compute
        node. Each line connected to the chipset represents a data path
        with a bandwidth equivalent to FSB1333.}
\end{figure}

The Woodcrest system at RRZE represents the prototypical design of
modern commodity HPC clusters: 217 compute nodes (see
Fig.~\ref{fig:woodcrest}) are connected to a single InfiniBand (IB)
switch (Voltaire ISR9288 with a maximum of 288 ports,
cf.~\cite{voltaire})\@. The dual-socket compute nodes (HP DL140G3) are
equipped with 8\,Gbytes of main memory, two Intel Xeon 5160 dual core
chips (codenamed ``Woodcrest'') running at 3.0\,GHz and the bandwidth
optimized ``Greencreek'' chipset (Intel 5000X)\@.
With Intel's new Core2 architecture several improvements were introduced
as compared to the Netburst design, aiming at higher instruction
throughut, shorter pipelines and faster caches to name a few which are
important for High Performance Computing.  Each node features a DDR IB
HCA in its PCIe-8x slot, thus the maximum IB communication bandwidth
(20\,GBit/s per direction at 10 bits per byte) is exactly matched to
the capabilities of the PCIe-8x slot (16 GBit/s per direction at 8
bits per byte)\@. The two-level IB switch is a compromise between DDR
and SDR (10 GBit/s per direction) technology: The 24 switches at the
first level which provide 12 downlinks to the compute nodes and 12
uplinks run at DDR speed, while the 12 second level switches run at SDR
speed. Thus, communication intensive applications can get a
performance hit when spread over several first level switches.

\subsection{Conroe -- InfiniBand cluster}

While multi-socket compute nodes are the standard building blocks of
HPC clusters they are obviously not the optimal choice for a wide
range of MPI-parallel HPC applications:
\begin{itemize}
\item A single network link must be shared by several sockets.
\item ccNUMA technology as used in, e.g., the SGI Altix or
	AMD Operon based systems bears the potential for performance
	penalties when locality contraints are not observed.
\item Bus overhead is introduced by cache coherency protocols.
\end{itemize}
Going back to the ``roots'' of cluster computing, as implemented by the single
socket Intel S3000PT board design, one can alleviate these problems: A
single socket is connected to one network link and to the local memory
through a single frontside bus (FSB1066)\@. While the
nominal bandwidth per socket is reduced as compared to the HP DL140G3
compute nodes (two FSB1333 connections for two sockets), power
consumption can be singnificantly improved through the use of
unbuffered DIMMs instead of fully buffered DIMMs. Note that the
power consumption of a single fully buffered DIMM can be as high as
5--10 Watts. Moreover, the lack of cache coherence traffic can overcome
the nominal loss in main memory bandwidth, resulting in an equal or
even higher main memory throughput per socket for the Conroe system as
measured with the TRIAD benchmark below.

The Xeon 3070 dual-core CPUs (codenamed ``Conroe'') used in this system 
implement the Intel Core2 architecture and run at 2.66\,GHz.
66 S3000PT nodes with 4\,Gbytes of memory each are connected to a 
72-port IB switch (Flextronics) running at full DDR speed.

\subsection{Sun UltraSPARC T2 -- single socket Sun server}\label{sec:N2}

The single socket ``Niagara 2'' system studied in this report is an early
access, pre-production model of Sun's T2 server series.
\begin{figure}[tbp]
\centerline{%
\includegraphics*[width=0.8\textwidth,angle=270]{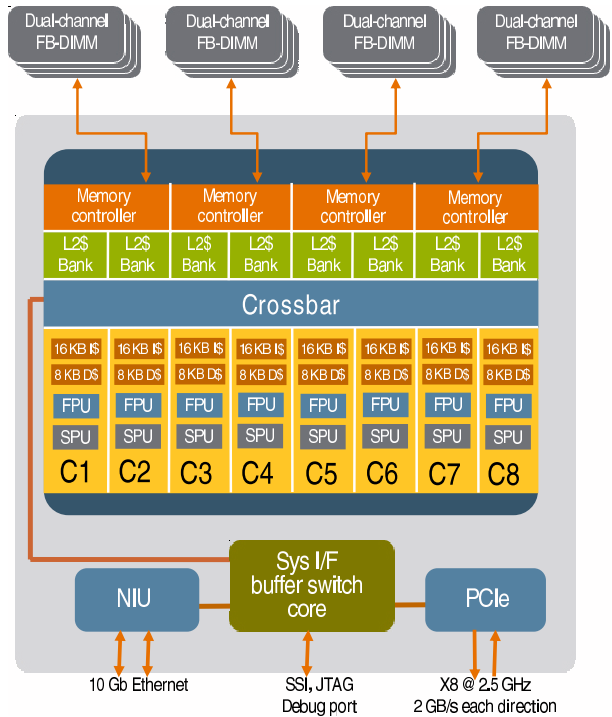}
}%
\caption{\label{fig:niagara-arch} Block diagram of the Sun UltraSPARC T2 (``Niagara 2'') 
chip architecture \cite{t2tech}. Eight physical cores ($C1,\ldots,C8$) with
local L1 data (8 KB) and L1 instruction (16 KB) caches are connected
to eight L2 banks (two of them sharing a memory controller) through a
non-blocking crossbar switch. Several interconnect ports (e.g.\ PCIe-8x
or 10\,Gb Ethernet) and a cryptographic coprocessor are put on the die,
complementing the ``server on a chip'' architecture. (Picture by courtesy 
of Sun Microsystems)}
\end{figure}
Trading high single core performance for a highly parallel system on a
chip architecture is the basic idea of Niagara 2 as can be seen in
Fig.~\ref{fig:niagara-arch}: Eight simple in-order cores (running at
1.4\,GHz) are connected to a shared,
banked L2 cache and four independently operating dual channel FB-DIMM
memory controllers through a non-blocking switch. 
At first glance the UMA memory subsystem provides
the scalability of ccNUMA approaches, taking the best of two worlds at
no cost. The aggregated nominal main memory bandwidth of 42\,Gbyte/s
(read) and 21\,Gbyte/s (write) for a single socket is far ahead of most
other general purpose CPUs and topped only by the NEC SX-8 vector
series. Since there is only a single floating point unit (performing
MULT or ADD operations) per core, the system balance of approximately 4
bytes/Flop (assuming read) is the same as for the NEC SX-8 vector
processor.

To overcome the restrictions of in-order architectures and long memory
latencies, each core is able to support up to eight threads. These threads
can be interleaved between the various pipeline stages with only few
restrictions \cite{t2tech}\@.  Thus, running more than a single thread per core is a
must for most applications.

Going beyond the requirements of the tests presented in this report
one should be aware that the Niagara 2 chip also comprises on-chip
PCIe-x8 and 10\,Gb Ethernet connections as well as a cryptographic
coprocessor.

\section{Low-level benchmarks and performance results}

\subsection{TRIAD}

TRIAD is based on the well-known vector triad code, which has been
extensively used
by Sch\"onauer~\cite{schoenauer} to quantify the capabilities of
memory interfaces. The triad performs a multiply-add
operation on four vectors, \texttt{A(:)=B(:)+C(:)*D(:)} in Fortran.
The loop kernel is repeated to produce accurate execution time and
cache performance measurements.

With its code balance of 2\,words/Flop, the vector triad is obviously
limited by memory bandwidth on all current supercomputer systems,
including vectors. If the arrays are short enough to fit into the
cache of a RISC processor, the benchmark tests the ability of the
cache to feed the arithmetic units. Even in this situation there is no
processor on which the triad is purely compute-bound. Consequently,
given the array lengths and basic machine performance numbers like
maximum cache and memory bandwidths and latencies, it should be easy
to calculate to highest possible performance of the vector triad.
Unfortunately, many of the currently popular PC-based systems fall
short of those expectations because their memory interface suffers 
from severe inefficiencies.

The usual write-back policy for outer-level caches leads to an
additional complication. As the cache can only communicate with memory
in chunks of the cache line size, a write miss kicks off a cache line
read first, giving the cache exclusive ownership of the line.  These
so-called RFO (read for ownership) transactions increase the code
balance even further to 2.5\,words/Flop. Some architectures support
special machine instructions to circumvent RFOs, either by bypassing
the cache altogether (``non-temporal'' or ``streaming'' stores on x86,
``block stores'' on Sun's UltraSPARC)
or by claiming cache line ownership without a prior read (``cache line
zero'' on IBM's Power architecture)\@. Often the compiler is able to
apply those instructions automatically if certain alignment constraints
are satisfied. It must be noted, though, that
cache bypass on write can have some impact on performance if the
problem is actually cache-bound.

While the vector triad code in RZBENCH is designed with MPI to allow
simple saturation measurements, this benchmark is most often used with
standard OpenMP parallelization. Unfortunately, OpenMP can have some
adverse effects on performance.  If, for instance, the applicability
of special instructions like non-temporal stores depends on the
correct alignment of data in memory, the compiler must ``play safe''
and generate code that can do without assumptions about alignment. At
best there will be two different versions of a loop which are selected
at runtime according to alignment constraints. If other restrictions
like, e.g., ccNUMA placement or load imbalance further confine 
available options, one can easily be left with large compromises
in performance. 

\begin{figure}[tbp]
\centerline{\includegraphics*[width=0.95\textwidth]{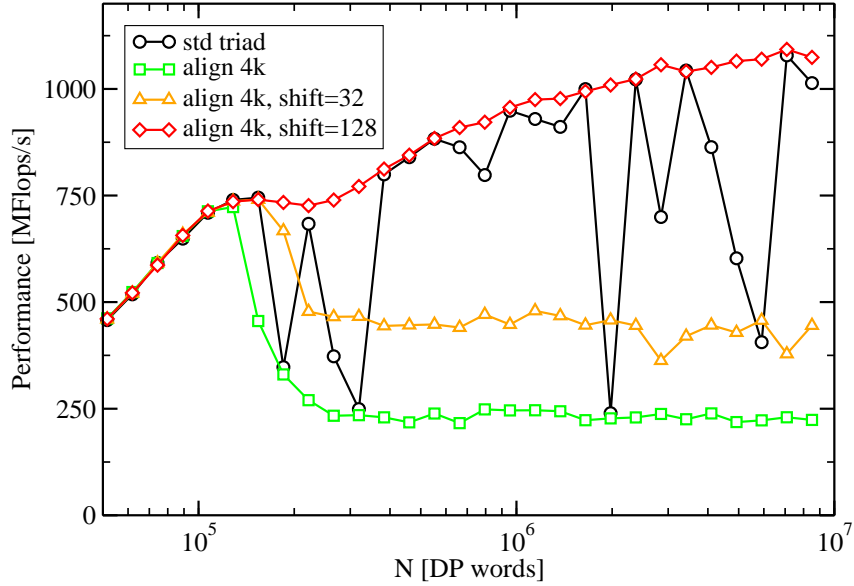}}%
\caption{\label{fig:N2seg}OpenMP-parallel vector triad performance versus
        array length for different alignment options on Sun UltraSPARC T2
        with 32 threads and static scheduling.}
\end{figure}
As an example we will consider the OpenMP vector triad on Sun's UltraSPARC T2
processor, described in Sect.~\ref{sec:N2}\@.  Without any special
provisions the vector triad performance with 32 threads shows a very
erratic pattern (circles in Fig.~\ref{fig:N2seg})\@. Threads were distributed
evenly across cores for these runs. Apparently, some
values for $N$ entail access patterns that utilize, in the worst case,
only one of the four available memory controllers at a time. This can be easily
explained by the fact that the controller to use is selected by
address bits 8 and 7, while bit 6 determines which of the two L2 banks 
to access~\cite{t2tech,hether}\@. If
$N$ is such that all threads always hit the same memory controller 
or even cache bank for all four data streams concurrently, performance 
breaks down by a factor of four. The typical ``lockstep'' access pattern emposed
by loop kernels that work on multiple data streams 
ensures this in a most reliable way if OpenMP chunk base
addresses are aligned inappropriately.
This condition can actually be enforced by manual alignment of 
\verb.A(:)., \verb.B(:)., \verb.C(:)., and \verb.D(:). to byte addresses
which are multiples of $512=2^9$\@. In Fig.~\ref{fig:N2seg}, the 
devastating effect of alignment to 4096 byte boundaries is shown
(squares)\@. 

Knowing the details about memory controller assigment, however, it is
easy to devise a mutual arrangement of arrays that avoids the
bottlenecks. After alignment to the 4\,kB boundary, the four arrays can
be shifted by four different integer multiples of some offset
$k$\@. The triangles and diamonds in Fig.~\ref{fig:N2seg} show the
results for $k=32$ and $k=128$, respectively. The latter case seems to
be optimal, which is not surprising since it constitutes the ``sweet
spot'' where all four controllers are addressed concurrently, independent
of $N$\@. All erratic behaviour has vanished.

It must be stressed that the Niagara 2 architecture shows a very rich
set of performance features, of which the influence of array alignment
is only one. Furthermore, the starting addresses for the 32 OpenMP
chunks that emerge from static scheduling have not been adjusted in
any special way. This may be insignificant on the Niagara 2, but it is
of vital importance on x86-based architectures where certain
vectorization instructions can only be applied for arrays that are
aligned to 16 byte boundaries. Details about if and how optimal
alignment and data placement can be achieved by special programming
techniques will be published elsewhere.

\subsection{IMB}

To test the basic capabilities of the interconnects we use the Intel
MPI benchmark (IMB 3.0) suite which is the successor of the famous Pallas MPI
suite.  Within this suite the standard benchmark to test the
unidirectional bandwidth and latency between two MPI processes is the
so called ``PingPong'': Using \verb.MPI_SEND./\verb.MPI_RECV. pairs a
message is sent from one to the other processor and upon arrival a
different message is sent back.
\begin{figure}[tbp]
\centerline{%
\includegraphics*[width=0.95\textwidth]{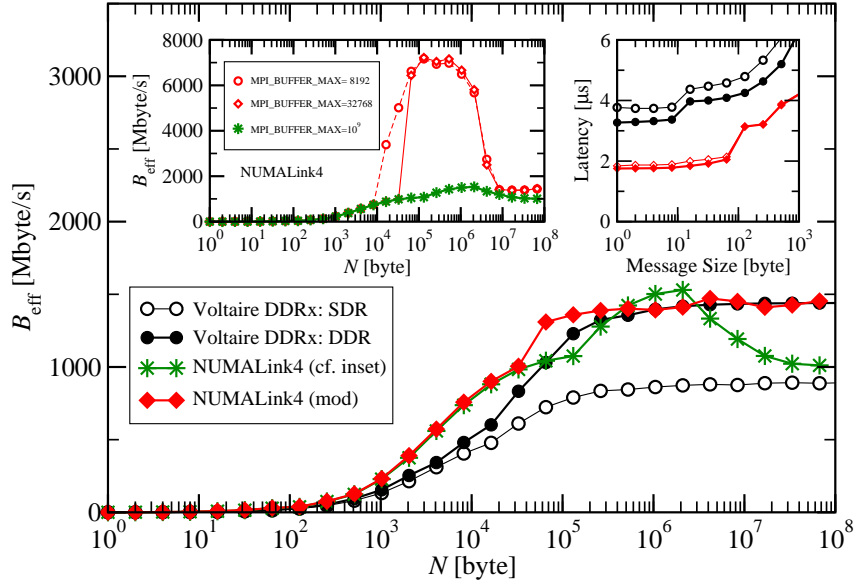}
}%
\caption{\label{fig:pingpong} MPI ``PingPong'' interconnect bandwith 
($B_\mathrm{eff}$) (main panel and left inset) and interconnect
latencies (right inset) as measured with the IMB\@. For SGI Altix,
bandwidth numbers of the standard IMB implementation and different values
of \texttt{MPI\_BUFFER\_MAX} are given in the left inset while in the
main panel results are included for a modified version (mod) of IMB,
which accounts for the shared-memory node architecture.}
\end{figure}
This is repeated a large number of times to get sufficient accuracy,
but it is important to note that the messages themselves are never 
touched, i.e.\ modified, in this scheme.

The main panel of Fig.~\ref{fig:pingpong} depicts the typical
unidirectional bandwidth vs.\ message size. The left inset shows
latency for the interconnects used in this performance
study. While the IB technologies behave the conventional way and
achieve approximately 70--75\% of their unidirectional bandwidth
limit, running the benchmark with no changes on the SGI Altix shows a
strange behaviour (left inset of Fig.~\ref{fig:pingpong})\@. A
bandwidth maximum of more than 7\,GB/s can be achieved at intermediate
message lengths, exceeding more than twice the nominal capabilities of
NUMALink4\@. For large messages performance breaks down to the IB DDR
level. 

Although results like this are readily interpreted by vendors to
show the superior quality of their products, a more thorough analysis
is in order. Keeping in mind that both processes involved in the
communication run in shared memory, the mystery is easily unraveled
(see Fig.~\ref{fig:pp}):
\begin{figure}[tbp]
\includegraphics*[width=0.43\textwidth]{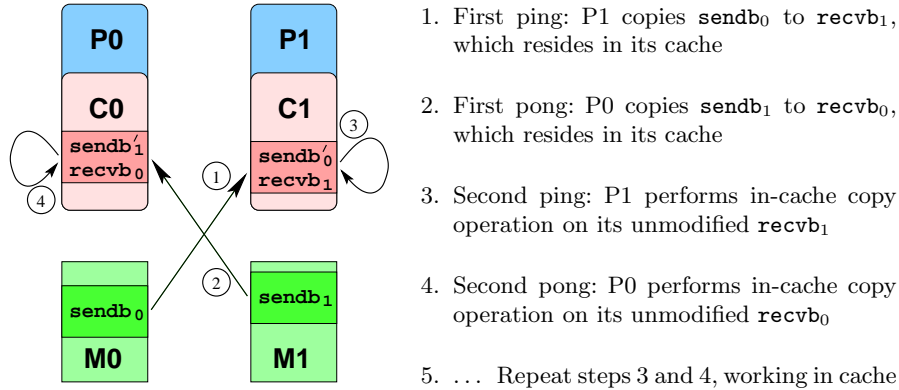}\hfill
\begin{minipage}[b]{0.55\textwidth}
\begin{enumerate}
\item First ping: P1 copies \verb.sendb.$_0$ to \verb.recvb.$_1$, which resides
        in its cache\bigskip
\item First pong: P0 copies \verb.sendb.$_1$ to \verb.recvb.$_0$, which resides
        in its cache\bigskip
\item Second ping: P1 performs in-cache copy operation on its unmodified \verb.recvb.$_1$\bigskip
\item Second pong: P0 performs in-cache copy operation on its unmodified \verb.recvb.$_0$\bigskip
\item \ldots~~Repeat steps 3 and 4, working in cache
\end{enumerate}
\end{minipage}
\caption{\label{fig:pp}Chain of events for the standard MPI PingPong
        on shared-memory systems when the messages fit in to cache. 
        C0 and C1 denote the caches of processors P0 and P1, respectively.
        M0 and M1 are P0's and P1's local memories.}
\end{figure}
The transfer of \verb.sendb.$_0$ from process 0 to \verb.recvb.$_1$ of
process 1 can be implemented as a single \emph{copy} operation on the receiver
side, i.e.\ process 1 executes \verb.recvb.$_1$\verb.(1:N). = \verb.sendb.$_0$\verb.(1:N)., where
\verb.N. is the number of bytes in the message. If \verb.N. is sufficiently 
small, the data from \verb.sendb.$_0$ is located in the cache of process 1 and there
is no need to replace or modify these cache entries unless
\verb.sendb.$_0$ gets modified. However the sendbuffers are not
changed on either process in the loop kernel. Thus, after the first
iteration the sendbuffers are located in the caches of the receiving
processes and in-cache copy operations occur in the succeeding
iterations instead of data transfer through the network.

There are two reasons for the performance drop at larger message
sizes: First, the L3 cache (9\,Mbyte) is to small to hold both or at
least one of the local receive buffer and the remote send
buffer. Second, the IMB is performed so that the number of repetitions
is decreased with increasing message size until only one iteration ---
which is the initial copy operation through the network --- is done for
large messages.

The use of single-copy transfers as described above can be controlled
on SGI Altix through the \verb.MPI_BUFFER_MAX. environment variable
which specifies the minimum size in bytes for which messages are
considered for single-copy.  As can be seen in the left inset of
Fig.~\ref{fig:pingpong}, changing the environment variable from its
default value 32768 one can adjust the width of the
artificial ultra-high bandwidth ``hump''\@. If
\verb.MPI_BUFFER_MAX. is larger than the biggest message, the effect 
vanishes completely. In this case, however, asymptotic performance
(stars in main panel of Fig.~\ref{fig:pingpong}) drops
significantly below the IB DDR numbers. This leads to the conclusion
that there is substantial overhead in this limit with single-copy
transfers disabled.

It is obvious that real-world applications can not make use of the
``performance hump''\@.  In order to evaluate the full potential
of NUMALink4 for codes that should benefit from single-copy for large
messages, we suggest a simple modification of the IMB PingPong benchmark: 
Adding a second ``PingPong'' operation in the inner iteration with arrays 
\verb.sendb.$_i$ and \verb.recvb.$_i$ interchanged (i.e.\ \verb.sendb.$_i$ 
is specified as the receive buffer with the second \verb.MPI_RECV. 
on process $i$), the sending process $i$ gains exclusive ownership of 
\verb.sendb.$_i$ again. After adjusting the timings in the code accordingly the
modified version shows the well known and sensible network
characteristics (diamonds in Fig.~\ref{fig:pingpong})\@.

In view of this discussion some maximum ``PingPong'' bandwidth numbers
for SGI Altix systems on the HPC Challenge website \cite{hpcchallenge}
should be reconsidered.

\section{Application benchmarks and performance results}

\subsection{Benchmark descriptions}

This section describes briefly the applications that constitute part
of the benchmark suite. We have selected the five most interesting
codes which were also used in the previous procurement by RRZE under
similar boundary conditions (processor numbers etc.)\@.

\subsubsection{EXX}

EXX is a quantum chemistry package developed at the chair for
theoretical chemistry at FAU\@. It is used for the calculation of
structural and electronic properties of periodic systems like solids,
slabs or wires, applying (time-dependent) Density Functional Theory
(DFT)\@. Performance is dominated by FFT operations using the
widely known FFTW package. The program is written in Fortran90 and
parallelized using MPI\@. 

The benchmark case contained in the suite is largely cache-bound and
scales reasonably well up to 32 cores. Note that EXX bears some
optimization potential (trigonometric function tabulation, FFT
acceleration by vendor libraries) which has been exploited by
benchmarking teams in the course of procurements. However, for
long-term reproducability and comparability of performance results the
codebase will not be changed to reflect architecture-specific
optimizations.

\subsubsection{AMBER/PMEMD}

AMBER is a widely used commercial molecular dynamics (MD) suite.
Dis\-tri\-bu\-ted-me\-mo\-ry FFT and force field calculations dominate
performance. The benchmark case ``HCD'' used for these tests simulates
HPr:CCpa tetramer dynamics using the \verb.PMEMD. module of AMBER\@.

This code is largely cache-bound but also suffers from slow networks.
The program is written in Fortran90 and parallelized with MPI\@.

\subsubsection{IMD}

IMD is a molecular dynamics package developed at the University
of Stuttgart. At FAU, it is mainly used by Materials Science groups
to simulate defect dynamics in solids. It is weakly dependent on
memory bandwidth and has moderate communication requirements.
The package was developed in C and is parallelized using MPI\@.
As the test case works with a $100^3$ lattice and the domain is decomposed
evenly in all three dimensions, 64 is the largest power-of-two
process number that can be used with it.

\subsubsection{Oak3D}

Oak3D is a physics code developed at the Institute for Theoretical
Physics at FAU\@.  It models the dynamics of exotic (superheavy)
nuclei via time-dependet Hartree-Fock (TDHF) methods and is used to
simulate photoabsorption, electron capture, nuclear fusion and
fission.  For calculating derivatives, the code relies heavily on
small-size FFTs that are usually handled inefficiently by
vendor-provided packages. This is why Oak3D uses its own FFT
package. Performance is dominated by FFT and dense matrix-vector
operations; for large processor numbers an \verb.MPI_ALLREDUCE. operation
gains importance. Some memory bandwidth is required, but benefits from large
caches can be expected. The code was developed with Fortran90 and
MPI\@.

\subsubsection{TRATS}

TRATS, also known as BEST, is a production-grade lattice-Boltzmann CFD
code developed at the Institute for Fluid Dynamics (FAU)\@. It is
heavily memory-bound on standard microcomputers, but compute-bound on
top of the line vector processors like the NEC SX because of its code
balance of $\approx 0.4$\,words/Flop.  It uses Fortran90 and MPI for
parallelization. Parallelization is done by domain decomposition, and
in the strong scaling benchmark case we chose a $128^3$ domain, cut
into quadratic slabs along the $x$-$y$ plane. While we are aware that
this is not an optimal domain decomposition, it allows us to
control the communcation vs.\ computation ratio quite easily. With
strong scaling it thus represents a powerful benchmark for network
capabilities.

TRATS is currently the
only code in the suite for which the execution infrastructure provides
a weak scaling mode, but this feature has not been used here.

We present performance results for application benchmarks in 
a concise format as it would be impossible to iterate over all
possible options to run the five codes on four architectures.
Dual core processors are used throughout, so we performed scalability
measurements by filling chips first and then sockets. That way, 
neighbouring MPI ranks share a dual-core chip when possible.
Strict process-core pinning (processor affinity) was implemented
in order to get reproducible and consistent results. On HLRB2,
all runs were performed inside a single standard (no high-density)
dedicated SSI node. The latest compiler releases available at the time of writing
were used (Intel 10.0.025 and Sun Studio 12)\@. For Niagara 2 only
a subset of the application benchmarks was considered. A more complete
investigation is underway.

\subsection{Performance results}

\subsubsection{Single core}

\begin{figure}[htbp]
\psset{yunit=3,xunit=3}
\centerline{%
\begin{pspicture}(-1.2,-1.1)(1.6,1.2)
  \plotaxesetc
  \plotdatafileopen{hlrb-b-1P.dat}{red}{o}
        \rput(1,-0.4){\psdot[dotstyle=o,dotscale=1.5,linecolor=red](0,0)}
        \rput[l](1.07,-0.4){HLRB2}
  \plotdatafileopen{woody-1P.dat}{blue}{square}
        \rput(1,-0.6){\psdot[dotstyle=square,dotscale=1.5,linecolor=blue](0,0)}
        \rput[l](1.07,-0.6){Woodcrest}
  \plotdatafileopen{pt-1P.dat}{green}{triangle}
        \rput(1,-0.8){\psdot[dotstyle=triangle,dotscale=1.5,linecolor=green](0,0)}
        \rput[l](1.07,-0.8){Conroe}
\end{pspicture}}
\caption{\label{fig:onecore}Single core performance comparison using the most
        important benchmarks from the RZBENCH suite. Numbers
        have been normalized to the best system for each benchmark.
	The parallel binary for PMEMD requires at least
	two MPI processes.}
\end{figure}
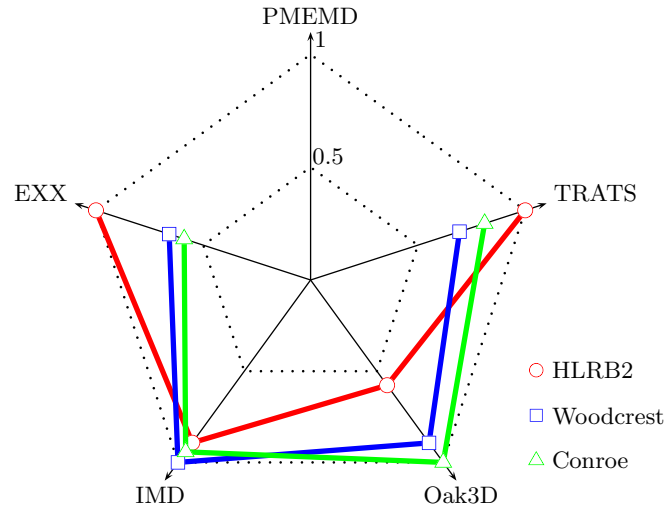

As a first step we compare the single core performance of HLRB2, the
RRZE Woodcrest cluster and the RRZE Conroe cluster in order to set
a baseline for scalability measurements. Extrapolating from raw
clock frequency and memory bandwidth numbers one might expect that 
Itanium cores could hardly be competitive, as measured by their
price tag. However, one should
add that the Itanium bus interface is much more efficient than on
Core2 in terms of achievable fraction of theoretical bandwidth,
even if the lack of non-temporal store instructions on IA64 is taken
into account.
Moreover, if the compiler is able to generate efficient EPIC code,
the IA64 architecture can deliver better performance per clock cycle
than the less clean, very complex Core2 design. Fig.~\ref{fig:onecore}
reflects those peculiarities. Interestingly, the largely cache-bound
codes EXX and AMBER/PMEMD (see Fig.~\ref{fig:onesocket} for PMEMD as the parallel
binary requires at least two MPI processes) that may be expected 
to scale roughly with clock frequency show superior performance on Itanium 2\@.
On the other hand, although Oak3D should benefit from large memory
bandwidth and big caches, it falls short of these expectations by roughly
40\,\%\@. This effect can be explained by the abundance of short loops
in the code which pose a severe problem for the in-order IA64
architecture.  Even if software pipelining can be applied by the
compiler, short loops lead to dominating wind-up and wind-down phases
which cannot be overlapped between adjacent loops without manual
intervention by hand-coded assembly~\cite{treibig}\@. Moreover,
latency cannot be hidden efficiently by prefetching. For the
lattice-Boltzmann code TRATS, however, IA64 is way ahead per core
because its memory architecture is able to sustain a large number of
cocurrent write streams. The results are thus in line with published
STREAM bandwidth numbers~\cite{stream}\@.

Interestingly the Conroe system, despite of its lower nominal
per-socket memory bandwidth (FSB1066) compared to Woodcrest (FSB1333),
outperforms the latter significantly on TRATS and Oak3D\@. Its simple
one-socket node design is obviously able to yield a much higher
fraction of theoretical peak bandwidth. For the cache-bound codes
IMD, EXX and AMBER Conroe suffers from its lower clock frequency.
We will see later that this can in some cases be overcompensated
by the superior per-socket network bisection bandwith of the Conroe
cluster.

\subsubsection{One and two sockets}

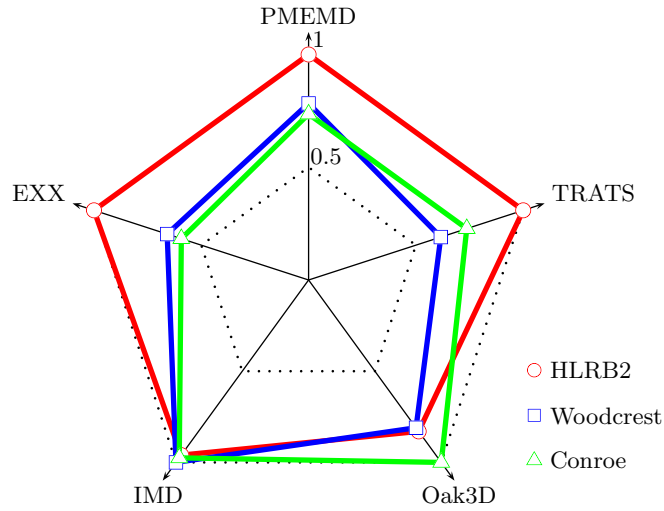
\begin{figure}[htbp]
\psset{yunit=3,xunit=3}
\centerline{%
\begin{pspicture}(-1.2,-1.1)(1.6,1.2)
  \plotaxesetc
  \plotdatafile{hlrb-b-1S.dat}{red}{o}
        \rput(1,-0.4){\psdot[dotstyle=o,dotscale=1.5,linecolor=red](0,0)}
        \rput[l](1.07,-0.4){HLRB2}
  \plotdatafile{woody-1S.dat}{blue}{square}
        \rput(1,-0.6){\psdot[dotstyle=square,dotscale=1.5,linecolor=blue](0,0)}
        \rput[l](1.07,-0.6){Woodcrest}
  \plotdatafile{pt-1S.dat}{green}{triangle}
        \rput(1,-0.8){\psdot[dotstyle=triangle,dotscale=1.5,linecolor=green](0,0)}
        \rput[l](1.07,-0.8){Conroe}
\end{pspicture}}
\caption{\label{fig:onesocket}Single socket performance comparison. In case
        of the Conroe system this is a complete node.
}
\end{figure}

The current evolution of multi-core processors shows the attractive
property that the price per raw CPU socket (or ``chip'') stays roughly constant over
time. If software can exploit the increased level of parallelism
provided by multi-core, this leads to a price/performance ratio that
follows Moore's Law. Of course, bottlenecks like shared caches, memory 
connections and network interfaces retard this effect so that it is vital to know
what level of performance can be expected from a single socket. This
data is shown in Fig.~\ref{fig:onesocket}\@.
Comparing with the single-core data in Fig.~\ref{fig:onecore}, the
most notable observation is that in contrast to the x86-based
processors the IA64 system is able to improve significantly on Oak3D
performance if the second core is used. This is mostly due to the
doubling of the aggregated cache size from 9\,MB to 18\,MB and because
two cores can sustain more outstanding references and
thus better hide latencies. For the other benchmarks,
scalability from one to two cores is roughly equivalent.

\begin{figure}[htbp]
\psset{yunit=3,xunit=3}
\centerline{%
\begin{pspicture}(-1.2,-1.1)(1.6,1.2)
  \plotaxesetc
  \plotdatafile{woody-2S.dat}{blue}{*}
        \rput(1,-0.4){\psdot[dotstyle=*,dotscale=1.5,linecolor=blue](0,0)}
        \rput[l](1.07,-0.4){2 sockets}
  \plotdatafile{woody-1S-2S.dat}{blue}{square*}
        \rput(1,-0.6){\psdot[dotstyle=square*,dotscale=1.5,linecolor=blue](0,0)}
        \rput[l](1.07,-0.6){1 socket}
  \plotdatafileopen{woody-1P-2S.dat}{blue}{pentagon*}
        \rput(1,-0.8){\psdot[dotstyle=pentagon*,dotscale=1.5,linecolor=blue](0,0)}
        \rput[l](1.07,-0.8){1 core}
\end{pspicture}}
\caption{\label{fig:woody1N}Scalability inside a two-socket Woodcrest node: Single
        core, single socket and two-socket performance. There is no one-core 
        data for PMEMD as the parallel binary for this benchmark needs 
	at least two MPI processes.}
\end{figure}
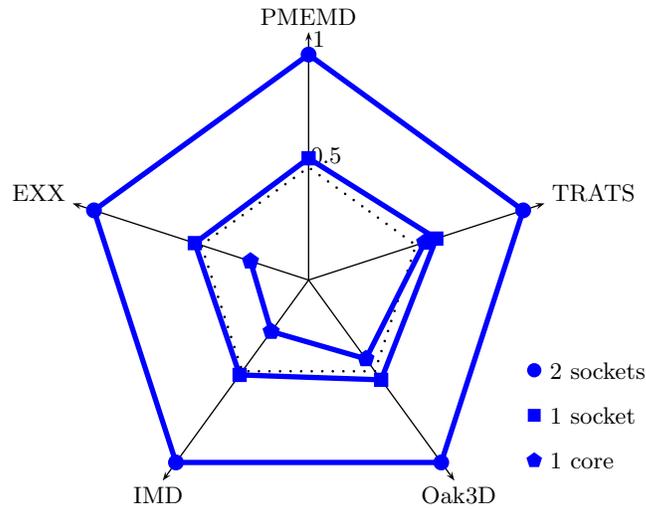

The two-socket Woodcrest nodes that are in wide use today deserve
some special attention here. Although the node layout suggests that
memory bandwidth should scale when using two sockets instead of one,
memory-bound benchmarks indicate that the gain is significantly
below 100\,\%\@. Fig.~\ref{fig:woody1N} shows a comparison between
one-core, two-core and four-core (two-socket) performance on
a single Woodcrest node. The cache-bound codes EXX and IMD
are obviously able to profit much better from the second core
on a chip than the bandwidth-limited Oak3D and TRATS\@. For Oak3D
this corroborates our statement that aggregate cache size
boosts two-core performance on HLRB2\@.

As suggested above, scalability of TRATS and Oak3D when going from one
to two sockets is less than perfect (improvements of 70\,\% and
80\,\%, respectively)\@. This result is matched by published STREAM
performance data~\cite{stream}\@. As for the reason one may speculate that
cache coherence traffic cuts on available bandwith although
the Intel 5000X chipset (``Greencreek'') has a snoop filter that
should eliminate redundant address snoops. In any case, we must emphasize
that in the era of multi-core processing it has become vital to understand
the topological properties and bottlenecks of HPC architectures and
to act accordingly by proper thread/process placement.

\subsubsection{Scalability}

In terms of scalability one may expect the SGI Altix system to
outperform the Intel-based clusters by far because of its NUMALink4
interconnect featuring 3.2\,GB/s per direction per socket.  However,
as mentioned above, even inside a single Altix node the network is not
completely non-blocking but provides a nominal bisectional bandwidth
of about 0.8\,GB/s per socket and direction only, which is roughly
equivalent to the achievable DDR InfiniBand PingPong performance using
a standard non-blocking switch. We thus expect scalability to show
roughly similar behaviour on all three architectures, certainly taking
into account the differences in single-socket performance.

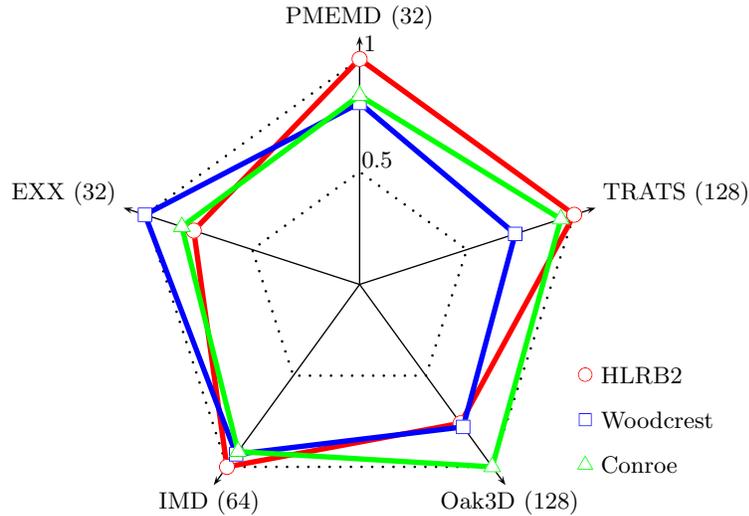
\begin{figure}[htbp]
\psset{yunit=3,xunit=3}
\centerline{%
\begin{pspicture}(-1.2,-1.1)(1.6,1.2)
  \plotaxesetcnum{PMEMD (32)}{EXX (32)}{IMD (64)}{Oak3D (128)}{TRATS (128)}
  \plotdatafile{hlrb-b-PAR.dat}{red}{o}
        \rput(1,-0.4){\psdot[dotstyle=o,dotscale=1.5,linecolor=red](0,0)}
        \rput[l](1.07,-0.4){HLRB2}
  \plotdatafile{woody-PAR.dat}{blue}{square}
        \rput(1,-0.6){\psdot[dotstyle=square,dotscale=1.5,linecolor=blue](0,0)}
        \rput[l](1.07,-0.6){Woodcrest}
  \plotdatafile{pt-PAR.dat}{green}{triangle}
        \rput(1,-0.8){\psdot[dotstyle=triangle,dotscale=1.5,linecolor=green](0,0)}
        \rput[l](1.07,-0.8){Conroe}
\end{pspicture}}
\caption{\label{fig:scaling}Parallel performance comparison. The number of MPI processes
        used is indicated for each benchmark.}
\end{figure}
Fig.~\ref{fig:scaling} shows a performance comparison for parallel
runs between 32 and 128 cores. The Conroe system can extend its lead
on Woodcrest especially for the network-bound parallel TRATS when
compared to the one-socket case (Fig.~\ref{fig:onesocket}) and even
draws level with HLRB2\@. This is due to its competitive network
bisection bandwidth. On PMEMD, despite its 10\,\% lower
clock frequency, Conroe can even slightly outperform Woodcrest
for the same reason.
In this context, EXX shows a somewhat atypical behaviour: Performance
on Woodcrest is far superior, despite the promising single core and single
socket data (Figs.~\ref{fig:onecore} and \ref{fig:onesocket})\@.
The reason for this is as yet unknown and will be investigated further.

\begin{figure}[htbp]
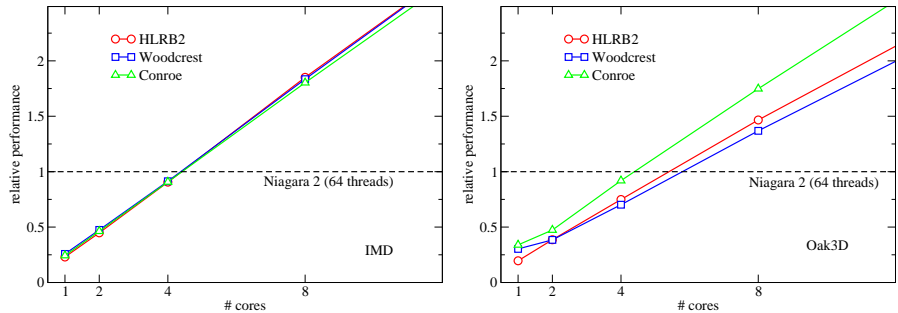

\noindent\includegraphics*[width=0.49\textwidth]{imd_rel.eps}\hfill
\includegraphics*[width=0.49\textwidth]{oak_rel.eps}
\caption{\label{fig:n2scaling}Comparing the Sun UltraSPARC T2 with
	the Intel-based architectures using IMD (left) and Oak3D (right).
	For reference, all data was normalized to the one-socket
	performance on Sun UltraSPARC T2 (dashed line).}
\end{figure}
Using the IMD and Oak3D benchmarks as prototypical cache and memory
bound codes, respectively, we finally compare the Intel-based 
architectures with Sun's Niagara 2 in Fig.~\ref{fig:n2scaling}\@.
Roughly, a single Niagara 2 socket is equivalent to between four
and six Intel cores. Note, however, that it takes 64 MPI processes
to reach this level of performance: Massive multithreading must make 
up for the rather simple design of the single core so that available
resources like memory bandwidth can be fully utilized and latencies can 
be hidden. 

\section{Conclusions and acknowledgements}

We have analyzed low-level and application benchmarks on current
high-performance architectures. On an early-access Sun Niagara 2
system we have shown that naive vector triad performance fluctuates
erratically with varying array size due to the hard-wired mapping of
addresses to memory controllers. Careful choice of alignment
constraints and appropriate padding allowed us to eliminate the
fluctuations completely, leading the way to architecture-specific
optimization approaches in the future. On HLRB2 we have explained a
widely unrecognized, pathological ``feature'' of the IMB Ping-Pong
benchmark and have shown a possible solution for making it more
meaningful for real applications.

RZBENCH, the integrated benchmark suite which has been developed by
RRZE, was then used to compare serial and parallel application
performance on HLRB2, a Woodcrest IB cluster and a Conroe IB cluster
with only one socket per node. The IA64 architecture shows superior
performance for most codes on a one-core and one-socket basis but is
on par with the commodity clusters for parallel runs.  It could be
shown that the extra investment in network hardware for a
single-socket commodity cluster can pay off for certain applications
due to improved bisection and aggregated memory bandwidth.
Sun's new UltraSPARC T2 processor could be demonstrated to display
very competitive performance levels if applications can sustain
a much more massive parallelism than on standard systems.

We are indebted to Sun Microsystems and RWTH Aachen Computing Centre
for granting access to a pre-production UltraSPARC T2 system.

\end{document}